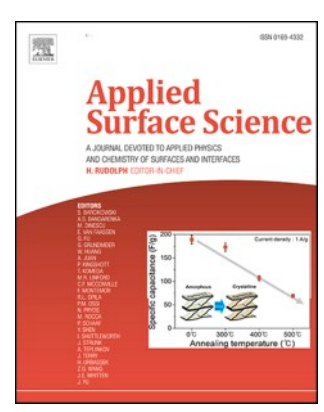

Full Length Article

# Pentagonal $PdTe_2$ monolayer for sustainable solar-driven hydrogen production

Narender Kumar [a,d], Shambhu Bhandari [b], Dario Alfè [b,c], Ravindra Pandey [d], Nacir Tit [a,*]

[a] Department of Physics, College of Science, UAE University, P.O. Box 15551, Al-Ain, United Arab Emirates
[b] Department of Earth Sciences and London Centre for Nanotechnology, University College London, Gower Street, London WC1E 6BT, United Kingdom
[c] Dipartimento di Fisica, Università di Napoli Federico II, Complesso Universitario di Monte Sant' Angelo – Via Cinthia, 21, 80126 Napoli, Italy
[d] Department of Physics, Michigan Technological University, 1400 Townsend Drive, Houghton, MI 49931, USA

ABSTRACT

This investigation demonstrates that the pentagonal $PdTe_2$ (penta-$PdTe_2$) monolayer is a highly tunable two-dimensional (2D) photocatalyst, characterized by the bandgap of 1.87 eV and high hole mobility. Using density functional theory calculations with the HSE06 functional, we show that tensile strain engineering (particularly at +2% and +3%) is essential for enabling spontaneous water splitting. At these strain values, the valence-band maximum and conduction-band maximum straddle the water redox potentials ($H+/H_2$ and $O_2/H_2O$) under both acidic (pH = 0) and neutral (pH = 7) conditions. The monolayer's low hole effective mass facilitates rapid charge extraction, mitigating recombination and driving the oxygen evolution reaction (OER) more effectively than many hexagonal and pentagonal counterparts. The Gibbs free energy (ΔG) pathways indicate that overpotentials for the hydrogen evolution reaction (HER) and OER are highly sensitive to mechanical deformation, specifically biaxial strain, through which +3% tensile strain, yielding an optimized balance of overpotentials of $\eta_{HER}$ = 0.70 V at pH = 0 and $\eta_{OER}$ = 0.72 V at pH = 7. Finally, integrating optical absorption with thermodynamic driving forces results in a Solar-to-Hydrogen (STH) efficiency of 20.40% at pH = 7. This exceeds the performance of several previously reported 2D catalysts, positioning penta-$PdTe_2$ as a superior candidate for sustainable, solar-driven hydrogen production.

## 1. Introduction

Pentagonal-based two-dimensional materials are considered a new emerging class of novel energy materials by possessing unique structural, electronic, optical and catalytic properties attributed mainly to their puckered and non-centrosymmetric morphology [1,2]. In particular, transition metal-based pentagonal 2D monolayers exhibit higher stability and semiconducting functionality, which are crucial for photocatalytic and water splitting applications [3–6]. However, among the many proposed structures, only a few pentagonal polymorphs have been experimentally synthesized [7,8], while many others remain hypothetical and await experimental validation. Recently, a pentagonal $PdTe_2$ (penta-$PdTe_2$) monolayer, which naturally crystallizes in the hexagonal 1 T structure in bulk form, has been synthesized using an epitaxially stabilized technique [8]. This synthesis was possible through a symmetry-driven epitaxial method by both direct tellurization of Pd (100) single-crystal surfaces at 500°C under ultrahigh vacuum and lattice matching with the square lattice of the substrate, which helped stabilize the pentagonal structure. The synthesized pentagonal $PdTe_2$ is characterized as an indirect semiconductor with an energy bandgap of 1.2 eV measured using scanning tunnelling spectroscopy (STS) and validated by electronic band structure calculations with a value of 1.05 eV using density-functional theory (DFT) with the Perdew-Burke-Ernzerhof (GGA-PBE) functional [8].

In recent years, several DFT-based studies on the pentagonal $PdTe_2$ have reported its several physicochemical properties. The monolayer possesses excellent thermoelectric properties, exhibiting a high thermal conductivity of up to 0.83 W $m^{-1}$ $K^{-1}$ [9]. The presence of Te vacancies induces both spin polarization and magnetic moments of magnitudes up to 1.87 $\mu_B$ and creates mid-gap states, modifying its electronic and optical properties [10]. Compressing the monolayer up to −10% leads to a semiconductor-to-metal transition, enabling switching ratios of $10^{11}$ for field-effect devices [11]. Moreover, the bilayer of pentagonal $PdTe_2$ exhibits a bulk photovoltaic effect with a shift current reaching 125.49 μA $V^{-2}$, which is attributed to spontaneous in-plane polarization of 1.70 × $10^{-10}$ C.m, significantly outperforming bilayer $PdSe_2$ and $PdS_2$ [12]. Very recently, Parkar et al. [13] demonstrated that Mo- and Ti-substituted pentagonal $PdTe_2$ exhibits excellent HER performance with reduced overpotentials down to 50 meV, comparable to Pt catalysts. However, this latter study was limited to the hydrogen evolution reaction (HER) half-reaction and did not address the oxygen evolution reaction (OER), which is equally critical for overall water splitting.

* Corresponding author.
*E-mail address:* ntit@uaeu.ac.ae (N. Tit).

Pentagonal 2D monolayers represent a fascinating and relatively recent frontier merging materials. Because a flat 2D crystalline plane cannot be tiled with regular pentagons without distortions, these materials naturally exhibit unique structural puckering, lower symmetry, and highly anisotropic properties [14]. Since the prediction of penta-graphene, the field has expanded to include binary and ternary pentagonal systems that are often more thermodynamically stable and synthetically viable, such as penta-Transition Metal Dichalcogenides (p-TMDs) (e.g., penta-$PdX_2$, with X = S, Se, Te and Janus derivatives) [3]. Unlike flat graphene, many p-TMDs are intrinsic semiconductors with indirect or direct bandgaps ranging from 1.0 eV to 2.5 eV, making them highly suitable for electronic devices without needing gap-engineering [3]. Because of the unique geometrical and physiochemical characteristics, they have been relevant for distinguished applications such as: (i) Photocatalytic water splitting: Many p-TMDs (like p-$PdS_2$ and p-$PdTe_2$) possess bandgaps that perfectly straddle the water reduction ($H^+$/H2) and oxidation ($H_2O/O_2$) potentials. The conduction-band minimum (CBM) must be more negative than the hydrogen reduction potential −4.44 eV vs vacuum at pH = 0; while the valence-band maximum (VBM) must be more positive than the oxygen oxidation potential −5.67 eV vs vacuum at pH = 0. Pentagonal $PdX_2$ inherently satisfy these requirements at neutral and acidic conditions without requiring external bias. Moreover, due to the structural asymmetry, electrons and holes often travel in different directions at vastly different speeds. This intrinsic spatial separation drastically suppresses charge recombination, boosting photocatalytic efficiency [15]. (ii) Energy Storage (Hydrogen and Li/Na Ions): Decorating the monolayer with light metals (e.g., Li, K, Na) can drastically enhance their adsorption energy for $H_2$ gas molecules [16]. (iii) Strain engineering: Due to their lower symmetry and buckled profiles, p-TMDs are incredibly sensitive to mechanical strain. Applying uniaxial of biaxial strain allows the tunning of bandgap from direct-to-indirect or even a semiconductor-to-metal transition. Applying strain can also alter the Gibbs free energy (ΔG) pathways of HER and OER to optimize catalytic performance [17]. Last but not the least, the use of noble metals like palladium (Pd) and platinum (Pt) is a cornerstone strategy in photocatalytic water splitting [18]. While they are too expensive to use as bulk materials, applying them as cocatalysts in microscopic or atomic amounts onto a host semiconductor (like $TiO_2$, g-$C_3N_4$, or 2D monolayers) drastically boosts solar-to-hydrogen (STH) conversion efficiency [18]. From all these perspectives, we selected the recently synthesized p-$PdTe_2$ monolayer for photocatalytic water splitting applications. In contrast to many other p-TMDs, p-$PdTe_2$ has not been explored in such applications while it has all potential characteristics for success as a competitor photocatalyst.

Furthermore, no prior work has systematically evaluated the band-edge alignment of pristine pentagonal $PdTe_2$ relative to water redox potentials at different pH values, nor has any prior work calculated the carrier effective masses and mobilities to assess charge transport efficiency or predicted solar-to-hydrogen (STH) conversion efficiencies. All of these are essential parameters for photocatalytic applications. To determine these parameters, we present a first-principles investigation of the photocatalytic water-splitting properties of pristine pentagonal $PdTe_2$ monolayers and examine improvements in efficiency via strain. We investigate both HER and OER processes at acidic (pH = 0) and neutral (pH = 7) environments, present the calculation of the complete Gibbs free energy pathways (ΔG) at equilibrium potential (U = 0), applied potential ($U = U_e$), and beyond the rate-determining step "RDS" (U > RDS) to determine overpotentials for full water splitting. We will calculate carrier effective masses (m*), deformation potentials ($E_d$, in eV), and in-plane elastic moduli ($C^{2D}$, in N/m) along both the x- and y-axis to assess anisotropic charge-transport properties. Band alignment analysis reveals whether the valence and conduction band edges straddle the water redox potentials ($H^+/H_2$ at 0 V and $O_2/H_2O$ at 1.23 V vs. Normal Hydrogen Electrode "NHE"), as a critical requirement for spontaneous photocatalytic water splitting without external bias. Additionally, we will evaluate the solar-to-hydrogen (STH) efficiency by integrating the optical absorption spectrum with the incorporation of thermodynamic driving forces, providing quantitative metrics for practical photocatalytic performance. This holistic approach will build on existing knowledge and propose the pentagonal $PdTe_2$ as a promising candidate for efficient photocatalytic water-splitting applications.

## 2. Methodology

First-principles calculations for the pentagonal $PdTe_2$ (penta-$PdTe_2$) monolayer were performed within density functional theory (DFT) using the Vienna *ab initio* simulation package (VASP) [19]. The electron–ion interaction was described using the projector augmented-wave (PAW) method, and the Kohn–Sham states were expanded in a plane-wave basis set with an energy cutoff of 550 eV. Exchange-correlation effects were treated using the generalized gradient approximation (GGA) [20]. In addition, the screened hybrid functional HSE06 [21] was employed to obtain improved electronic band-gap estimates. For HSE06 calculations, the Hartree-Fock exchange fraction was set to 0.25 (AEXX = 0.25), and the screening parameter was set to 0.2 Å$^{-1}$ (HFSCREEN = 0.2). The optimized primitive cell of penta-$PdTe_2$ contains two palladium (Pd) atoms and four tellurium (Te) atoms arranged in the Cairo-tessellated arrangement. Periodic-image interactions along the out-of-plane direction were minimized by inserting a vacuum spacing of 20 Å along the *z*-axis. Long-range dispersion interactions were included via Grimme's DFT-D3 correction [22]. Brillouin-zone integrations were carried out using a Monkhorst–Pack 9 × 9 × 1 k-point mesh [23] for geometry relaxation, and a denser grid of 12 × 12 × 1 k-point mesh for density of states calculations. Geometry optimizations were considered converged when the total-energy change was below $10^{-6}$ eV and the residual forces on each atom were less than 0.01 eV/Å. For the photocatalytic assessment, overall water splitting was evaluated under acidic and neutral conditions, corresponding to pH = 0 and pH = 7, respectively. The Gibbs free-energy changes (ΔG) for the HER and OER pathways were computed following the procedure described in the Electronic Supplementary Information (ESI), Equations S1-S17. The pH-dependent redox potentials were determined using the Nernst equation [24], according to which the water redox potentials shifted by 0.059 × *pH* unit at 298.15 K (Eq. S8-S9). The dynamical stability of the penta-$PdTe_2$ monolayer was examined by performing phonon dispersion calculations using PHON code [25]. In this method, small atomic displacements are introduced into a periodically repeated supercell, and the resulting Hellmann-Feynman forces are used to construct the interatomic force-constant matrix. For the present system, a 5 × 5 × 1 supercell of the optimized penta-$PdTe_2$ monolayer was employed to obtain well-converged phonon frequencies.

## 3. Results and discussion

### 3.1. Structural and electronic properties

The pristine penta-$PdTe_2$ monolayer is a Cairo-tessellated arrangement of Palladium (Pd) and Tellurium (Te) atoms [8]. The optimized primitive cell (top and side view) of penta-$PdTe_2$ is shown in Fig. 1. Unlike the conventional hexagonal $PdTe_2$ monolayer of $P\overline{3}m1$ space group, the penta-$PdTe_2$ monolayer exhibits monoclinic $P2_1/c$ space group with lower symmetry [8,26]. As shown in Fig. 1, each Pd atom is tetra-coordinated, forming four bonds with Te atoms; two from the upper and two from the lower layer, resulting in a Te–Te dimerization across the layer that gives rise to the puckered morphology. The average bond lengths between Pd-Te and Te-Te are 2.62 Å and 2.80 Å, respectively. The projected in-plane atomic configuration shows a well-defined rectangular unit cell with the calculated lattice parameters, $a = 6.03$Å, $b = 6.37$Å and $\alpha = \beta = \gamma = 90^{\circ}$, which are very close to experimental and previous theoretical studies [5,8,10,11,27]. The thickness of the monolayer ($h = 1.79$ Å) is considerably shorter than that of the hexagonal $PdTe_2$ ($h = 2.76$ Å) monolayer [27]. It is worth noting that the pentagonal monolayer shares the same $C_2h$ symmetry as the previously synthesized penta-$PdX_2$ (X = S, Se) monolayers [7,28].

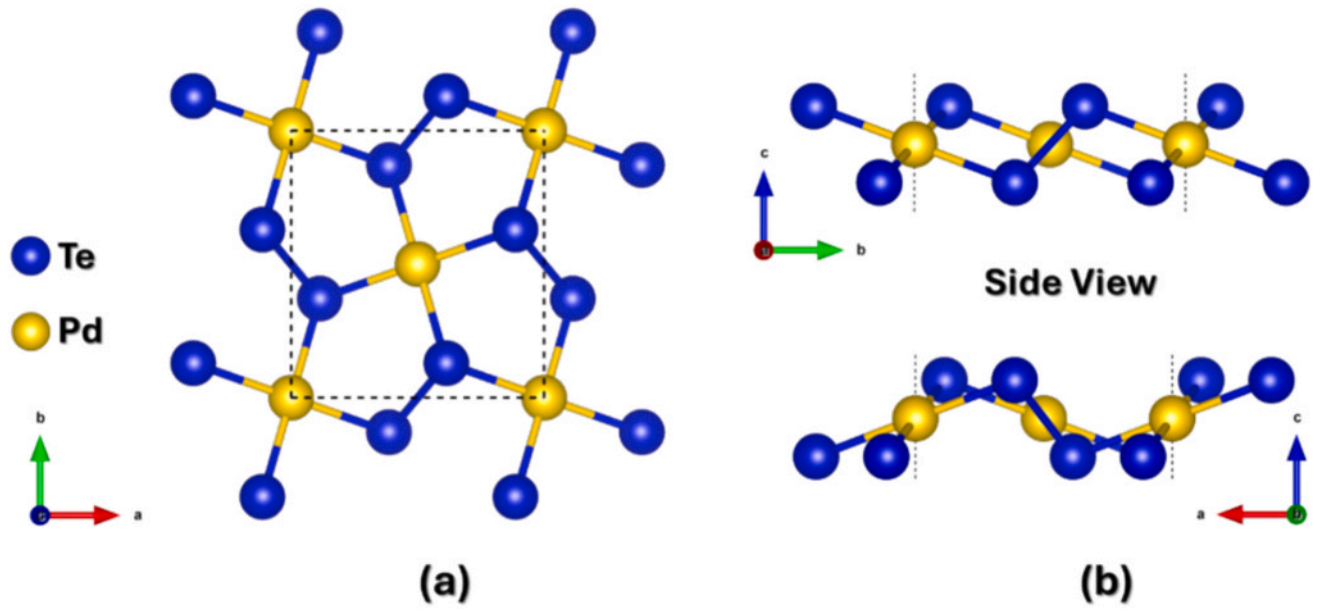

**Fig. 1.** Top and side views of the optimized structure of the pristine pentagonal $PdTe_2$ monolayer.

The dynamical and thermal stability of the penta-$PdTe_2$ monolayer was examined using phonon dispersion and ab-initio molecular dynamics (AIMD) simulations, as shown in Fig. 2. The phonon spectrum in Fig. 2a shows no negative frequencies throughout the high-symmetry lines of the Brillouin zone, confirming the dynamical stability of the optimized monolayer. To further evaluate its thermal robustness, AIMD simulations were performed at 300, 500, and 800 K. As presented in Fig. 2b, the total energy fluctuates with an amplitude less than 1% of the total energy value during time interval up to 8000 fs without any sudden energy drift, indicating that the structure remains thermally stable even at elevated temperature (i.e. up to 800 K). The relaxed atomic configurations after AIMD simulations, shown in Fig. 2c, further confirm that the penta-$PdTe_2$ monolayer is preserved at all the considered temperatures, with no obvious bond breaking or structural reconstruction.

To gain deeper insight into charge redistribution within the pristine penta-$PdTe_2$ monolayer, we have plotted the charge density difference (CDD) in three dimensions (Fig. 3a) and its planar-average projection along the z-axis (Fig. 3b), accompanied by a Bader charge analysis. The yellow and cyan regions correspond to charge accumulation and depletion, respectively. Pronounced charge accumulation around Pd atoms and depletion near Te atoms indicate a net charge transfer from Te to Pd, quantified as approximately +0.32 e per Te atom and −0.65 e per Pd atom from the Bader charge analysis. Although the electronegativity difference between Pd ($\chi = 2.20$) and Te ($\chi = 2.10$) is relatively small. However, the charge transfer might arise from strong hybridization between Pd-*4d* and Te-*5p* orbitals, as well as from the structural asymmetry of the puckered lattice. This hybridization leads to a mixed metallic-covalent bonding character and stabilizes the Cairo-tessellated network of penta-$PdTe_2$ monolayers.

The planar-averaged charge density profile $\rho(z)$ (Fig. 3b) further supports this observation, showing alternating peaks of electron accumulation and depletion near the Te sublayers. The asymmetry of the charge distribution implies partial out-of-plane polarization, consistent with the puckered geometry, which leads to shorter interlayer vertical distance $h = 1.79$Å as compared to $h = 2.76$Å of a hexagonal $PdTe_2$ monolayer [27].

Thereafter, we calculated the projected electronic band structure and corresponding total/projected density of states (TDOS/PDOS) of the pristine penta-$PdTe_2$ monolayer shown in Fig. 4. The band dispersion (Fig. 4a) shows that the penta-$PdTe_2$ monolayer exhibits indirect semiconducting bandgap behavior with VBM and CBM located between $Y-\Gamma$ and $\Gamma-X$ high-symmetry lines, respectively. Since DFT calculations with the PBE exchange–correlation functional underestimate the band gap, the band gap values obtained from the hybrid HSE06 functional are used for photocatalytic study. The calculated HSE06 band gap of 1.87 eV is in agreement with the previously reported HSE06-level theoretical values [3,5,10,11]. However, this value is larger than the experimentally reported scanning tunnelling spectroscopy (STS) band gap of ~1.2 eV for penta-$PdTe_2$ monolayer grown on a metallic Pd(100) substrate [8]. The difference can be attributed to the different physical conditions considered: the experimental STS value corresponds to a substrate-supported monolayer, where substrate screening, interfacial coupling and tunneling to the metallic Pd substrate can modify the apparent band gap; whereas the present calculation considers a free-standing monolayer. Analysis of PDOS (Fig. 4b) reveals that the VBM is primarily dominated by Te 5p orbitals (blue), whereas the CBM mainly originates from Pd 4d states. The strong hybridization between Pd-*4d* and Te-5*p* orbitals near the band edges confirms the covalent character of the Pd–Te bonding. The symmetric TDOS of spin up and spin down states further confirms the paramagnetic semiconducting character of the penta-$PdTe_2$ monolayer. Furthermore, the influence of spin–orbit coupling (SOC) on the band structure was evaluated using the HSE06 + SOC (Fig. S1, ESI) and found to have a negligible effect of order of −0.03 eV.

It is well known that mechanical strain can play an important role,

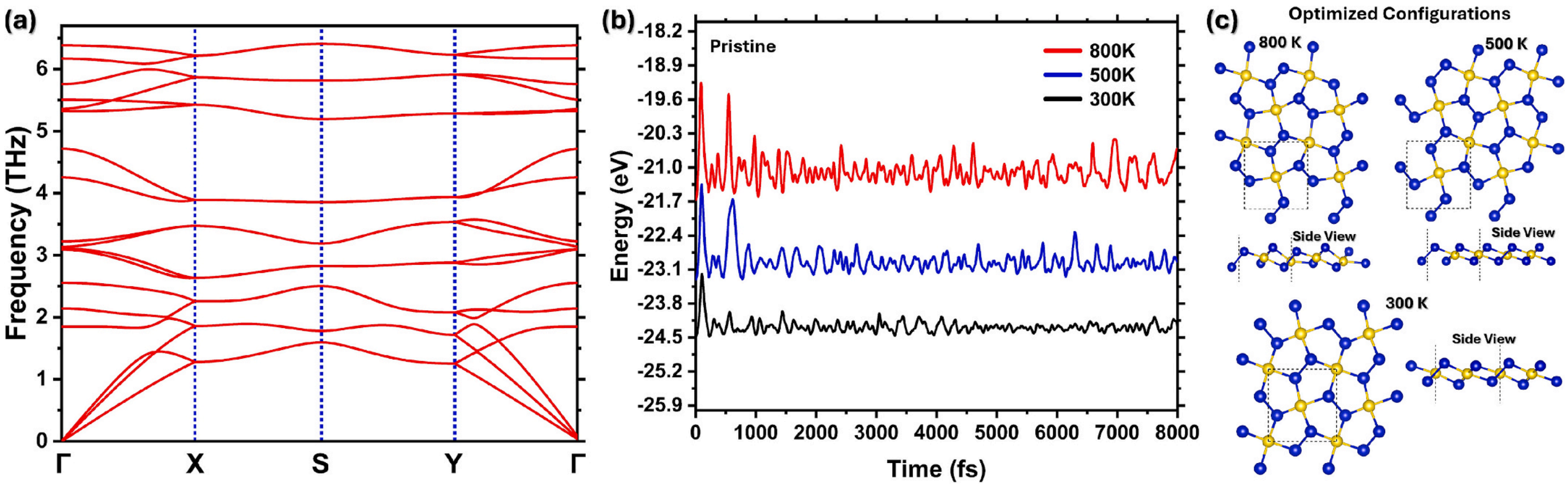

**Fig. 2.** (a) Phonon dispersion spectra of pristine penta-$PdTe_2$ monolayer. (b) Total energy evolution using ab-initio molecular dynamics (AIMD) simulations at 300, 500, and 800 K. The energy curves at 500 and 800 K are vertically shifted by 1.5 and 3.5 eV, respectively, for visual representation. (c) Relaxed configurations of penta-$PdTe_2$ monolayer after AIMD simulations at 300 K, 500 K and 800 K.

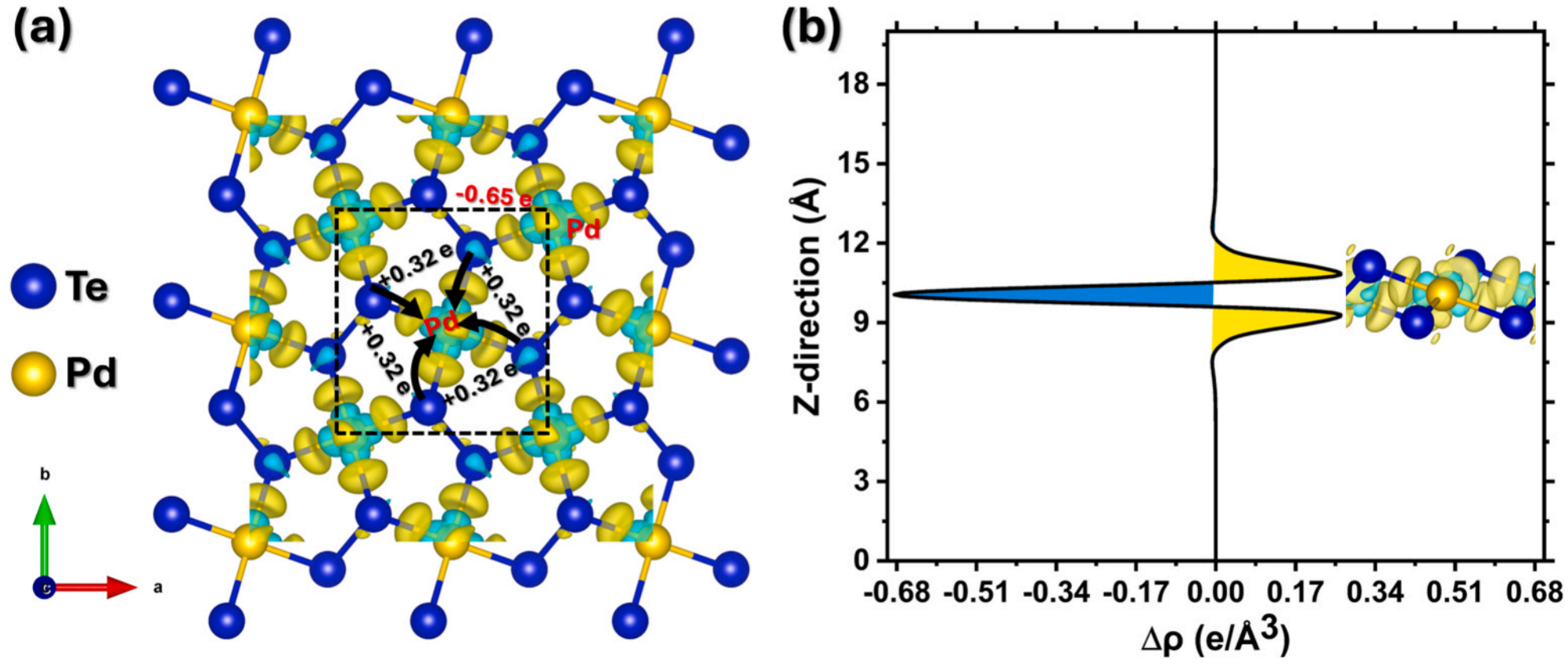

**Fig. 3.** Pentagonal $PdTe_2$ monolayer: (a) The charge density difference (CDD) plot, and (b) planar-average projection along the z-axis ρ(z). The *iso*-surface value is 0.005 e/Å$^3$.

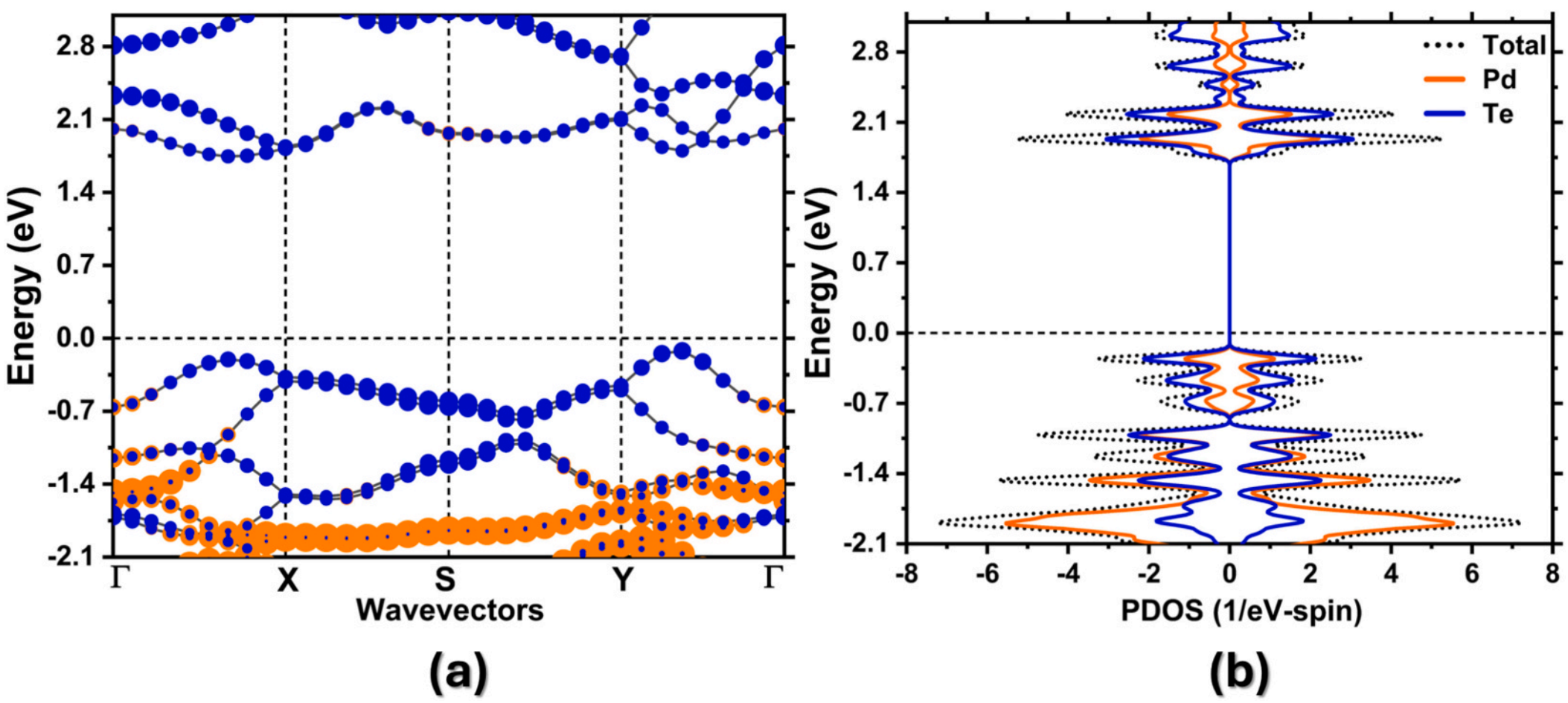

**Fig. 4.** Projected band structure and density of states at the HSE06 level theory. Color code: orange and blue for Pd and Te atoms, respectively.

and based on this, one can develop strategies to modulate the electronic structure of 2D materials, enabling systematic control over band dispersion, carrier transport, optical absorption, and catalytic performance. Therefore, uniform biaxial strain ($\varepsilon_{xy}$) ranging from −4% (compressive) to +4% (tensile) was applied on the penta-$PdTe_2$ monolayer to investigate strain-induced band modifications and their implications on overall water splitting.

Fig. S2 (ESI) summarizes the calculated HSE06 band structures under compressive (−4% to −2%) and tensile (+2% to +4%) biaxial strain. The band gap remains indirect under these biaxial strain ranges, except that a tensile +2% strain results in a direct band gap nature. Under the compressive strain (Fig. S2a-b, ESI), both VBM and CBM move downward in energy, accompanied by a gradual relocation of CBM. For instance, the calculated bandgap energies of 1.76 eV, 1.69 eV and 1.63 eV correspond to the compressive strains of −2%, −3% and −4%, respectively. This reduction in the gap is primarily driven by enhanced orbital overlap resulting from lattice contraction, which destabilizes the conduction-band states. However, under tensile strain (Fig. S2c-d, ESI), the lattice expansion reduces orbital hybridization, pushing the VBM upward while shifting the CBM to higher energies. Consequently, the bandgap increases, for instance, to 1.90 eV at a tensile strain of +2%, where the monolayer exhibits a direct semiconductor nature. At +3% and +4% strain, the VBM shifts closer to the X–S high-symmetry line, indicating a reordering of the valence-band states and reduction of the energy gap to 1.77 eV (+3%) and 1.63 eV (+4%), while the indirect bandgap character is maintained.

### 3.2. Band alignment and carrier mobility

To further elucidate the photocatalytic capability of penta-$PdTe_2$ monolayer for overall water splitting, the band edge positions were calculated with respect to the vacuum level and compared to the redox potentials for the hydrogen evolution reaction (HER, $E_{H^+/H_2}$) and oxygen evolution reaction (OER, $E_{O_2/H_2O}$) at acidic (pH = 0) and natural (pH = 7) medium. For an efficient photocatalyst, the CBM must lie above the hydrogen reduction potential, while the VBM should lie below the oxygen oxidation potential of water. As shown in Fig. 5, both the CBM and VBM of penta-$PdTe_2$ monolayer straddle the standard redox potentials of water at pH = 7 only (dotted red lines), confirming its thermodynamic suitability for driving both HER and OER under visible-light excitation. However, at pH = 0 (dotted blue line) only the CBM lies above the redox potential of hydrogen (−4.44 eV at pH = 0), similar to the behavior reported for the hexagonal β-$PdTe_2$ monolayer [29]. This result indicates that the HER can take place, but the OER (half-reaction) cannot proceed simultaneously in the case of a strongly acidic medium.

In addition, we also investigated how biaxial strain influences the photocatalytic suitability of the penta-$PdTe_2$ monolayer. Fig. 5 illustrates the shifts in CBM and VBM under biaxial strain ranging from −4% (compressive) to +4% (tensile). Under compressive strain, both band edges shift downward, slightly improving VBM alignment at pH = 7 but still failing to span the full water-redox window. In contrast, tensile strain induces a more favorable downward shift in both the CBM and VBM. Notably, at +2% and +3% tensile strain, the band edges successfully encompass both the oxidation and reduction potentials at pH = 0 and pH = 7, whereas at +4% tensile strain, the CBM shifts beyond the required threshold at pH = 0. Therefore, +2% and +3% biaxial tensile strains are identified as the most stable and optimal regimes for enabling overall water splitting. Among these, +3% biaxial strain is selected as the representative condition for the subsequent photocatalytic reaction analysis because it provides stronger strain-induced band-edge modulation while maintaining suitable redox alignment at both pH = 0 and pH = 7.

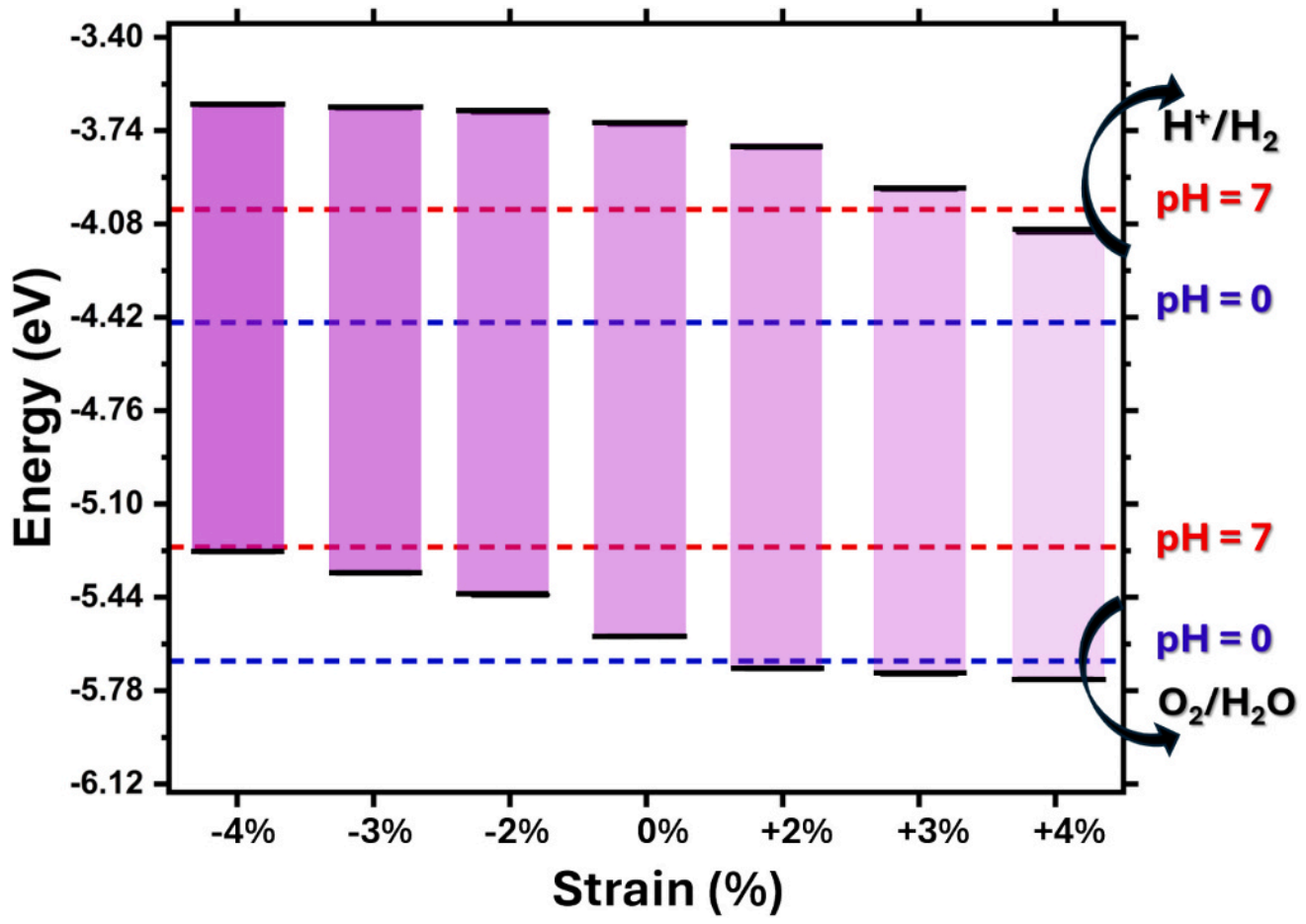

**Fig. 5.** Band gap alignment of pristine penta-$PdTe_2$ with respect to applied biaxial strain. The dotted blue and red lines represent the reduction and oxidation potentials at pH 0 and 7, respectively.

Along with band alignment, carrier mobility ($\mu_c$) is an important quantity for assessing the transport dynamics of photoexcited charge carriers. The $\mu_c$ is evaluated using the effective-mass approximation in conjunction with the 2D elastic modulus ($C_{2D}$) under the framework of the deformation potential theory (DPT) [30]. This approach inherently accounts for carrier scattering by long-wavelength acoustic phonons, the dominant scattering mechanism at room temperature, and thus provides an upper bound on the intrinsic carrier mobility. Furthermore, the anisotropy in carrier transport arising from the x- or y-direction of the penta-$PdTe_2$ monolayer, as well as variations in the effective mass and elastic properties, is effectively captured with this method. The $C_{2D}$ is calculated as:

$$C_{2D} = \frac{1}{S_0}\left(\frac{\partial^2 E_{\text{total}}}{\partial \varepsilon^2}\right) \tag{1}$$

where $S_0$ is the equilibrium area of the unstrained unit cell, $E_{\text{total}}$ is the total strain energy and $\varepsilon$ is the applied monoaxial strain as shown in **Figure-S3a, ESI**. The calculated $C_{2D}$ of 22.77 $N/m$ and 51.32 $N/m$ along the x- and y-axis, monoaxial strains are in good agreement with previously reported studies [5,24]. The $C_{2D}$ of the penta-$PdTe_2$ monolayer along the x and y axes is lower than that of penta-$PdX_2$ (X = S, Se), whereas it is higher compared to hexagonal-$PdTe_2$ monolayers [29,31] as shown in Table 1. A higher $C_{2D}$ indicates a mechanically more rigid lattice that is less susceptible to strain fluctuations, thereby reducing phonon-induced scattering and enhancing carrier mobility.

Furthermore, the $\mu_c$ for 2D penta-$PdTe_2$ monolayer is expressed as:

$$\mu_c = \frac{e\hbar^3 C_{2D}}{k_B T m^* \overline{m} E_d^2} \tag{2}$$

where $e$, $\hbar$, $C_{2D}$, $T$, $k_B$, $\overline{m}$, $m^*$, $E_d$ and $\mu_c$ denote the electron charge, reduced Planck's constant, in-plane elastic modulus, absolute temperature (300 K), Boltzmann constant, average effective mass ($\overline{m} = \sqrt{m_x^* m_y^*}$), effective mass along the transport direction, deformation potential constant, and the calculated carrier mobility, respectively. Here, effective mass ($m^*$) is defined as the inverse proportional to the curvature of the electronic band near the band edge and can be expressed as:

$$m^* = \pm\hbar^2\left(\frac{d^2E_k}{dk^2}\right)^{-1} \tag{3}$$

where $E$ and $k$ denote the energy and wavevector, respectively. The calculated carrier effective masses ($m^*/m_0$) of the penta-$PdTe_2$ monolayer are 1.47 and 1.32 along the x- and y-directions for electrons and 0.38 and 0.34 for holes, respectively. Compared to penta-$PdS_2$ (0.87/0.25) and penta-$PdSe_2$ (1.88/0.49), the electron effective masses of penta-$PdTe_2$ are 69 times higher along the x-axis than $PdS_2$, while 46 times lower than penta-$PdSe_2$ monolayer. Conversely, along the y-axis, they are significantly higher than penta-$PdS_2$ (by ~428 times) and ~169 times higher than penta-$PdSe_2$. For holes, the effective masses of penta-$PdTe_2$ are 47–81% smaller than those of $PdS_2$ and 58–77% smaller than those of $PdSe_2$, indicating lighter hole transport in the Te-based system. In comparison with the hexagonal-$PdTe_2$ monolayers (Table 1), the penta-$PdTe_2$ monolayer possesses significantly higher electron effective masses along both x and y-directions, suggesting higher carrier mobility and anisotropy in pentagonal monolayers. The results of calculations of carrier mobility along the x and y directions are shown in Table 2 and are compared with existing literature data.

Next, the deformation-potential constant ($E_d$) is calculated as $E_d = \frac{\partial E_{\text{edge}}}{\partial \varepsilon_{x/y}}$, where $E_{\text{edge}}$ denotes the shift of the VBM or CBM induced by applying uniaxial strain $\varepsilon = \Delta l / l_0$ along the $x$ or $y$ direction. Note that the deformation potential constant governs the scattering rate associated with electron-acoustic phonon interactions. Therefore, smaller values of the deformation potential constant should correspond to weaker carrier-phonon coupling and, consequently, higher carrier mobility. As shown in Fig. S3b, ESI the deformation potential is obtained by fitting the strain-dependent variation of the band-edge energies. All band-edge energies $E_{\text{edge}}$ are taken from electronic band-structure calculations performed at the HSE06 level of theory. The Fermi level is used as the reference when determining the deformation potential $E_d$. For penta-$PdTe_2$, the electron deformation potential are 1.78 eV and 1.62 eV along the x and y-directions, respectively, which are significantly smaller than those of penta-$PdX_2$ (X = S, Se) monolayers (Table 1). While the deformation potential of holes is somewhat larger, 2.17 eV along the x-axis and 4.53 eV along the y-axis, indicating stronger hole-phonon coupling along the y-direction. In comparison with hexagonal $PdTe_2$ monolayers [31], the distinct behaviour of the pentagonal lattice is found. The electron $E_d$ of hexagonal β-$PdTe_2$ (2.70/3.38 eV) and hexagonal-$PdTe_2$ (1.30/3.74 eV) are generally found larger along the y-direction, suggesting stronger electron–phonon coupling in the hexagonal structures. Using all these quantities, we finally calculated carrier mobility and found the electrons mobility to be 47.22 $cm^2.V^{-1}.s^{-1}$ and 159.35 $cm^2.V^{-1}\cdot s^{-1}$ along x and y direction, respectively. Whereas the hole mobilities are significantly higher, 475.47 $cm^2\cdot V^{-1}\cdot s^{-1}$ (along x-axis) and 307.17 $cm^2\cdot V^{-1}\cdot s^{-1}$ (along y-axis). This clearly indicates that holes are the primary high-mobility carriers in penta-$PdTe_2$, especially along the x-direction like other penta-$PdX_2$ (X = S, Se) monolayers. This behaviour is consistent with the relatively small hole effective masses and moderate deformation potentials reported earlier, which together reduce phonon-limited scattering and enhance hole transport. In comparison with hexagonal $PdTe_2$ monolayers, pentagonal $PdTe_2$ displays a very different transport character. For β-$PdTe_2$ (hexagonal), the electron mobility is extremely high along x (1258.47 $cm^2\cdot V^{-1}\cdot s^{-1}$) but very low along y (16.71 $cm^2\cdot V^{-1}\cdot s^{-1}$), while holes show the opposite anisotropy (107.92 *vs* 544.86 $cm^2\cdot V^{-1}\cdot s^{-1}$ for $x/y$). The other hexagonal-$PdTe_2$ monolayer shows exceptionally high electron mobilities in both directions (2066.32 and 848.16 $cm^2\cdot V^{-1}\cdot s^{-1}$), whereas hole mobilities are relatively poor ($\approx 25-28$ $cm^2\cdot V^{-1}\cdot s^{-1}$). These results imply that hexagonal polymorphs are strongly electron-dominated, with very high n-type mobility but relatively suppressed hole transport. In the context of water splitting, the high hole mobility of penta-$PdTe_2$ is beneficial for rapid extraction of photogenerated holes toward OER sites, whereas the hexagonal $PdTe_2$ phase is more suitable as an electron-transport component in reduction-dominated configurations. In addition, the carrier relaxation time ($\tau = \mu m^*/e$) was estimated at 300 K, and the corresponding electron relaxation time is found to be about 39.47 fs (119.61 fs) and for holes relaxation time to be 102.74 fs (59.39 fs) along x-axis (y-axis), respectively. The holes had longer relaxation time along x-axis while the opposite trend for electrons.

**Table 1**
Calculated effective mass ($m^*$), deformation $E_d$ (eV) and in plane elastic modulus $C_{2D}$ (N/m) along x and y-axis and their comparison with β-$PdTe_2$ (hexagonal) and hexagonal-$PdTe_2$ and other pentagonal materials.

| Material | Carrier type | $\frac{m^*}{m_0}(x)$ | $\frac{m^*}{m_0}(y)$ | $E_d(x)$ | $E_d(y)$ | $C_{2D}(x)$ | $C_{2D}(y)$ | Ref. |
|---|---|---|---|---|---|---|---|---|
| Penta-$PdTe_2$ | e | 1.47 | 1.32 | 1.78 | 1.62 | 22.77 | 51.32 | This work |
| | h | 0.38 | 0.34 | 2.17 | 4.53 | | | |
| Penta-$PdS_2$ | e | 0.87 | 0.25 | 8.59 | 9.40 | 58.00 | 82.00 | [32] |
| | h | 0.72 | 1.75 | 2.12 | 3.11 | | | |
| Penta-$PdSe_2$ | e | 1.88 | 0.49 | 5.24 | 5.03 | 38.44 | 98.18 | [33] |
| | h | 0.90 | 1.45 | 1.12 | 1.89 | | | |
| β-$PdTe_2$ | e | 0.27 | 1.56 | 2.70 | 3.38 | 40.55 | 8.89 | [29] |
| (hexagonal) | h | 0.84 | 0.32 | 4.34 | 1.48 | | | |
| hexagonal-$PdTe_2$ | e | 0.62 | 0.21 | 1.30 | 3.74 | 36.66 | 42.18 | [31] |
| | h | 0.66 | 0.84 | 7.54 | 7.53 | | | |

**Table 2**
Calculated carrier mobility $\mu_c(cm^2V^{-1}S^{-1})$ along x/y-axis, and calculated overpotential for HER ($\eta_{HER}$) and OER ($\eta_{OER}$) and their comparison with hexagonal $PdTe_2$ and other pentagonal materials at pH = 0 and U = 0.

| Material | Carrier type | $\mu_c(x)(cm^2V^{-1}S^{-1})$ | $\mu_c(y)(cm^2V^{-1}S^{-1})$ | Ref. |
|---|---|---|---|---|
| Penta-$PdTe_2$ | e | 47.22 | 159.35 | This work |
| | h | 475.47 | 307.17 | |
| Penta-$PdS_2$ | e | 40.97 | 169.11 | [32] |
| | h | 339.25 | 91.73 | |
| Penta-$PdSe_2$ | e | 29.40 | 229.01 | [33] |
| | h | 534.55 | 184.59 | |
| β-$PdTe_2$ (hexagonal) | e | 1258.47 | 16.71 | [29] |
| | h | 107.92 | 544.86 | |
| hexagonal-$PdTe_2$ | e | 2066.32 | 848.16 | [31] |
| | h | 27.94 | 25.33 | |

### 3.3. Photocatalytic properties of Penta-$PdTe_2$ monolayer

To evaluate the thermodynamic feasibility of photocatalytic water splitting on the penta-$PdTe_2$ monolayer, we systematically investigated the Gibbs free-energy (ΔG) profiles of both the HER and OER using the computational hydrogen electrode (CHE) framework. Solvation effects were explicitly included to capture the solid–liquid interfacial thermodynamics (see ESI section for Methods). The external potentials of photogenerated charge carriers directly affect the photocatalytic activity and are calculated from previous literature [34,35]. At pH = 0 (acidic medium), the photogenerated electrons ($U_e$) and photogenerated holes ($U_h$) values of 0.72 and 1.14 eV are obtained, respectively. The hydrogen reduction potential shifts as function of pH as $\Delta G_{pH} = -k_B T \ln(10) \times pH = -0.059 \times pH$ leading to a decrease in $U_e$ and an increase in $U_h$ with increasing pH. Accordingly, at pH = 7 (neutral medium), $U_e$ decreases to 0.31 eV, while $U_h$ increases to 1.55 eV, confirming sufficient redox driving forces under neutral conditions.

The HER proceeds via a two-step Volmer-Heyrovsky mechanism. In the first step, the adsorption of a proton-electron pair leads to the formation of an *H intermediate, while the second step involves the combination of *H with another proton and electron to release an $H_2$ molecule. To identify the most favorable adsorption configuration for the reaction intermediates, all possible adsorption sites were systematically examined. As shown in Fig. S4 (ESI), six distinct sites (S1–S6) were considered and their calculated total energies and $E_{ads}$ are presented in Table S1, ESI. Among these, the *H intermediate initially placed at sites S4 and S5 migrates spontaneously toward the top of the Te atom (site S2) exhibits the highest energetic active site with a binding energy of $E_{ads}$ = 1.15 eV at an optimal adsorption height of ~1.71 Å. Bader charge analysis further indicates that the adsorbed *H species donates an average charge of −0.72$e$ to the penta-$PdTe_2$ surface, confirming the occurrence of charge transfer during the adsorption process. The calculated HER Gibbs free energy change ($\Delta G_{HER}$) under acidic (pH = 0) and neutral (pH = 7) conditions are shown in Fig. 6a and b. At zero applied potential (U = 0), formation of the *H intermediate constitutes the rate-limiting step (RDS), with an unfavorable $\Delta G_{HER}$ of 1.24 eV (pH = 0) and 1.65 eV (pH = 7). Although the subsequent $H_2$ formation step is exothermic by nature, the intrinsic photogenerated electron potential is insufficient to fully compensate this barrier (red line). Consequently, an additional external electron potential ($\eta_{HER}$) of 0.53 eV at pH = 0 and 1.25 eV at pH = 7 or higher than the RDS (blue line) is required to effectively trigger the HER pathway thermodynamically downhill. Compared with other well-known 2D pentagon catalysts such as; $PdS_2$ (1.175 eV) [33], $PdSe_2$ (1.22 eV) [36], $HgS_2$ (0.65–1.66 eV) [37], $PtS_2$ (1.28 eV) [38] and the 2H phase $MoS_2$ (~2.0 eV) [39], the penta-$PdTe_2$ monolayer has comparable $\Delta G_{HER}$, which indicates that the penta-$PdTe_2$ has potential application prospects in the field of photocatalytic water decomposition.

Furthermore, the OER on penta-$PdTe_2$ follows the conventional four-electron pathway involving *OH, *O, and *OOH intermediates. In the first step, a water molecule is oxidized to form the *OH intermediate. This is followed by the deprotonation of *OH to yield *O intermediate. In the third step, the *O species interacts with an additional $H_2O$ molecule to generate the *OOH intermediate. Finally, oxidation of *OOH releases an oxygen molecule, completing the catalytic cycle. The $E_{ads}$ of all intermediates (*OH, *O, *OOH) are presented in Table S1, ESI. To identify the most stable adsorption configurations, several initial geometries were examined for the larger oxygenated intermediates, particularly *OH and *OOH. These included tilted, O-facing and H-facing orientations toward penta-$PdTe_2$ monolayer. During structural relaxation, the titled or H-facing configurations were spontaneously reoriented, making the O atom moving toward the surface to form the adsorbate–surface bond. This indicates that O-mediated adsorption is energetically preferred for these intermediates. Therefore, the lowest-energy relaxed O-bound configurations were used for the $E_{ads}$ and ΔG calculations. The results indicate that *OH preferentially binds at the Te-top site (S2 site) with an adsorption energy of ~1.41 eV; initial configurations placed at alternative sites (S1 and S5) relax spontaneously to the S2 site. The optimized O–H bond length is 0.99 Å, with an adsorption height of 2.07 Å above the surface. The *O intermediate exhibits stronger binding at the Pd–Te bridge site (S3), with an adsorption energy of ~1.90 eV, while configurations initially placed at the Pd site (S1) also migrate to this location. By contrast, *OOH adsorption is significantly less favourable, with a minimum adsorption energy of ~4.58–4.59 eV, reflecting the intrinsic difficulty of O-O bond formation. The optimized *OOH geometry exhibits an O–O–H bond angle of 101.0° and an adsorption height of ~2.20 Å. Consistent with these trends, Bader charge analysis reveals substantial charge transfer from the oxygenated intermediates to the substrate: −0.66 e for *OH, −1.24 e for *O, and −0.55 e for *OOH. The Gibbs free energy profile (Fig. 6c and d) reveals that the deprotonation of *OH to form *O is the rate-determining step of the OER, exhibiting the largest free-energy barrier of 3.22 eV at pH = 0 and 2.82 eV at pH = 7, respectively. Even under the intrinsic photogenerated hole potential ($U_h$), all OER steps remain uphill, indicating that water oxidation cannot proceed spontaneously. Notably, at pH = 7, the first and third steps become downhill under the influence of $U_{h;}$ however, the second and fourth steps remain energetically unfavourable. Consequently, additional external hole potential ($\eta_{OER}$) of 2.09 eV at pH = 0 and 1.18 eV at pH = 7 or higher than RDS (blue line) is required for all OER intermediates to become downhill. With increasing pH, the OER barrier decreases, and the required external potential to trigger the oxidation half-reaction decreases. This trend indicates that OER proceeds more favorably in a neutral medium than in acidic environments.

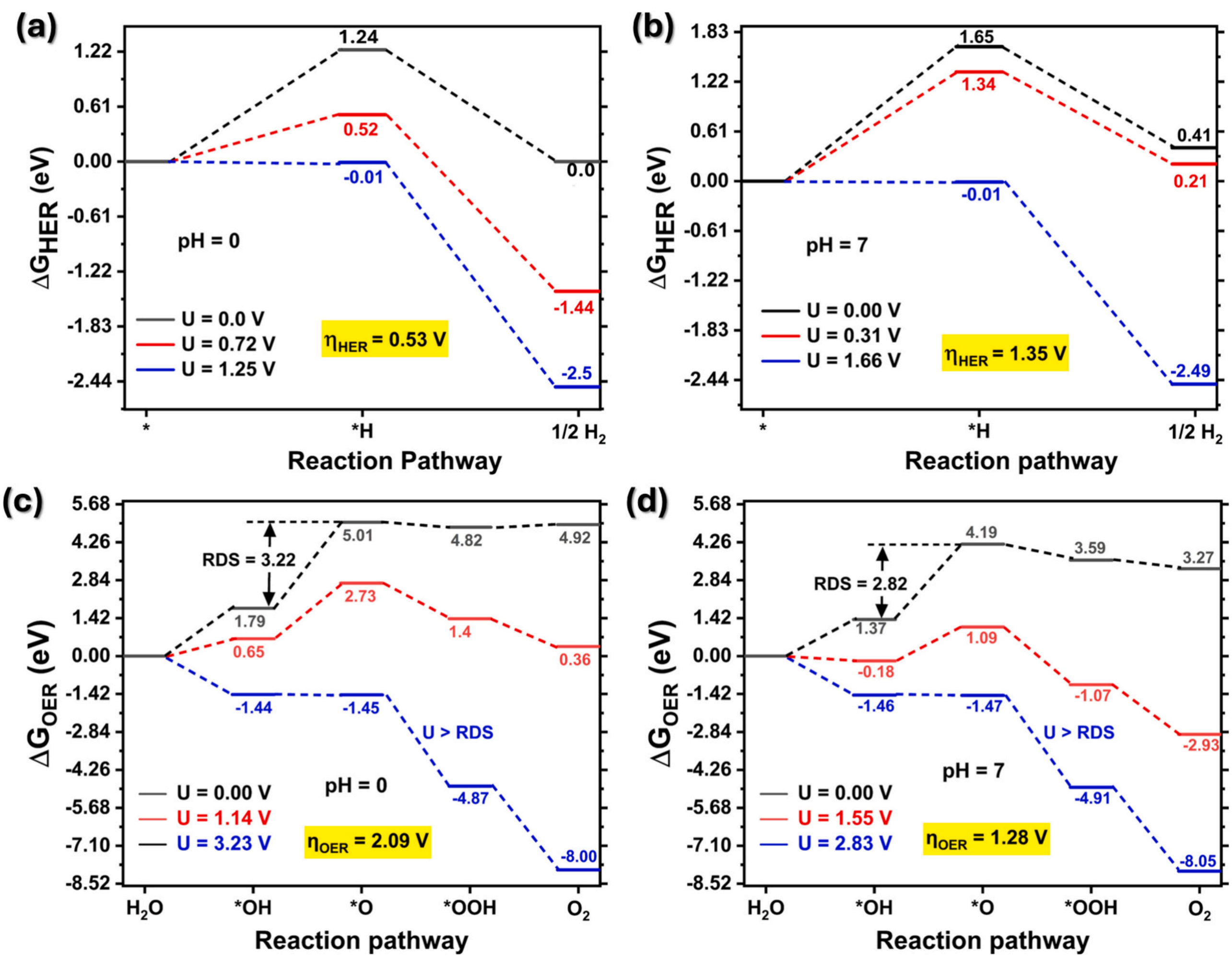

**Fig. 6.** Calculated free energy change for HER at potential U = 0 (black line), $U = U_e$ (red line) and U > RDS (blue line) at a) pH = 0, b) pH = 7, and OER at c) pH = 0, d) pH = 7.

It is well known that strain can tune the electronic structure of 2D photocatalysts (as discussed in the previous section), consequently, their redox driving force and surface reaction energies [40–43]. In penta-$PdTe_2$, biaxial deformation modifies the band-edge positions relative to vacuum, thereby directly affecting the thermodynamic feasibility of HER and OER at different pH conditions. We recall that Fig. 5 summarizes the strain-dependent band-edge alignment of penta-$PdTe_2$ computed at the HSE06 level, showing that tensile strain systematically shifts the CBM/VBM and alters the available photovoltage for reduction and oxidation half-reactions.

Motivated by the favorable band-edge alignment at moderate tensile strain, we focus on +3% biaxial strain as a representative condition to evaluate the reaction thermodynamics. This strain condition was chosen because it provides clear strain-induced modulation of the electronic structure while maintaining suitable CBM and VBM alignment with the $H^+/H_2$ and $O_2/H_2O$ redox potentials at both pH = 0 and pH = 7. To further confirm the structural stability of the +3% strained monolayer, AIMD simulations were performed at 300 K, 500 K and 800 K, as shown in Fig. S5, ESI. The total energy fluctuates within an amplitude of about ~1 eV of less than 1% of total energy value during the simulation time interval (up to 8000 fs), indicating that the +3% biaxially strained penta-$PdTe_2$ monolayer remains thermally stable even at high temperatures (500 K and 800 K). The final relaxed configurations also preserve the pentagonal $PdTe_2$ monolayer framework without significant structural distortion or bond breaking. These results confirm that the selected +3% biaxial strain condition is structurally feasible for subsequent photocatalytic reaction analysis. The corresponding ΔG for HER and OER at both pH = 0 and pH = 7 are shown in Fig. 7. At U = 0, the formation of the *H intermediate remains the rate-determining step, with $\Delta G_{HER}$ of 1.18 eV at pH = 0 and 1.60 eV at pH = 7 (Fig. 7a,b). These barriers are significantly lower than those of the unstrained penta-$PdTe_2$ monolayer. Under +3% applied strain, the intrinsic photogenerated electron potentials are $U_e = 0.49V$ at pH = 0 and $U_e = 0.07V$ at pH = 7. Note that the $U_e$ at pH = 7 is negligible; therefore, the minimum required potential for water decomposition (1.23 eV) is used as the reference case. The intrinsic $U_e$ partially compensates the HER barrier but does not fully eliminate it. Only when an external potential exceeding the rate-determining step (U > RDS) is applied, all HER steps become downhill. These results indicate that tensile strain enhances HER activity by weakening *H binding, thereby promoting a more optimal adsorption regime. Furthermore, the OER exhibits a more pronounced strain response. As shown in Fig. 7(c,d), the rate-determining step under +3% tensile strain shifts to the **O to *OOH* transformation, with free-energy barriers of 2.81 eV at pH = 0 and 2.40 eV at pH = 7 at $U = 0$. Compared with the unstrained system, the application of tensile strain stabilizes the *O intermediate, thereby lowering the barrier associated with *OH deprotonation. Under the influence of the intrinsic photogenerated $U_h$ (red lines), the *O intermediate and the subsequent $O_2$ formation step are further stabilized, particularly in neutral conditions. Nevertheless, at $U_h$ alone, not all OER steps become downhill. A sufficiently large external hole potential is still required to overcome the rate-determining barrier, yielding minimum required overpotentials of $\eta_{OER} = 1.54$ V at pH = 0 and 0.72 V at pH = 7 for all steps to proceed thermodynamically downhill.

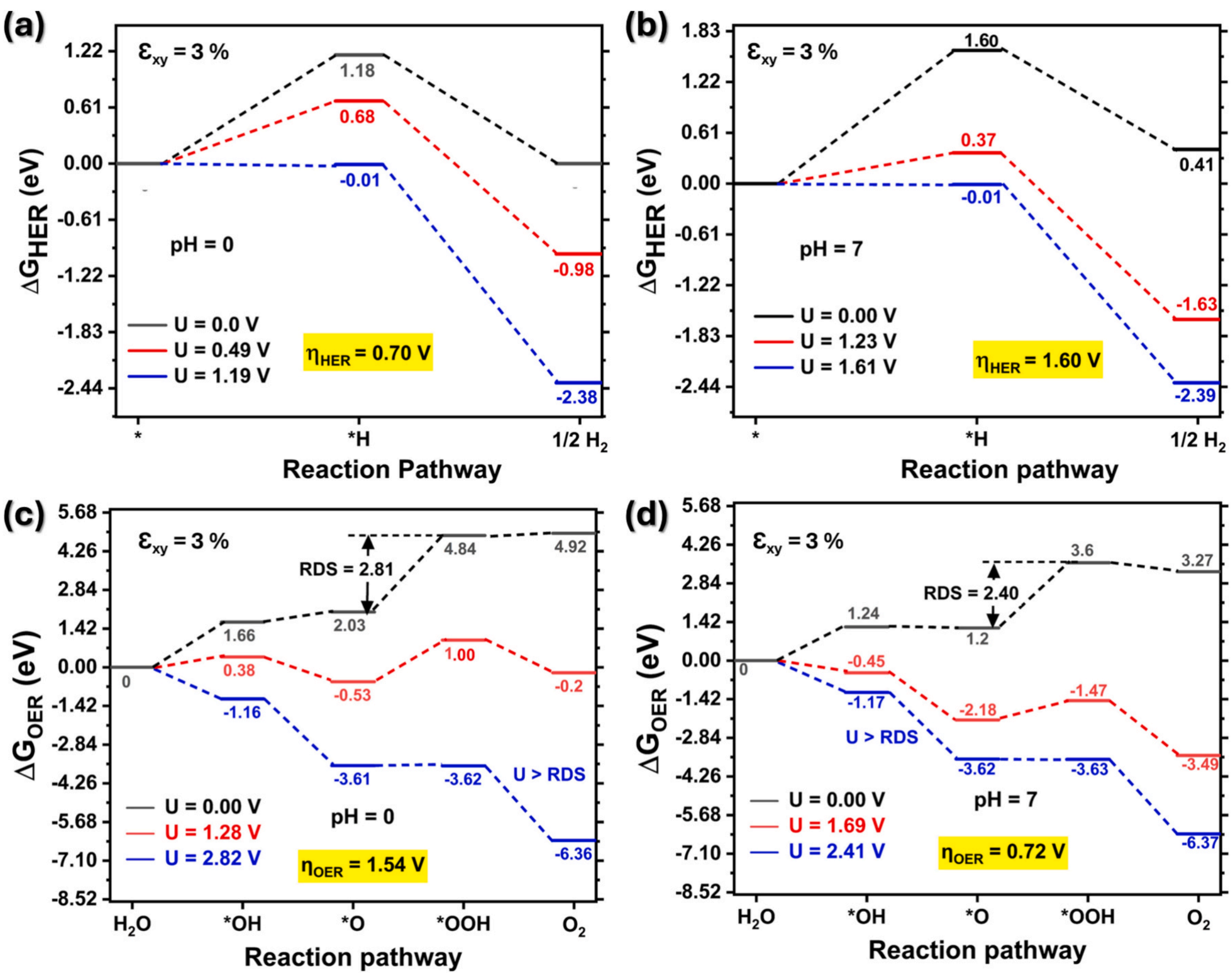

**Fig. 7.** Calculated ΔG for the HER at U = 0 (black), U = $U_e$(red), and U > RDS(blue) at (a) pH = 0 and (b) pH = 7, and for the OER at (c) pH = 0 and (d) pH = 7 under +3% biaxial strain.

### 3.4. d-band center

To understand the electronic origin of the (photo)electrocatalytic activity of penta-$PdTe_2$ and the strain effect, we have evaluated the d-band center ($\varepsilon_d$) of the active site. The hybridization between Pd *d* states and adsorbate (*H, *O, *OH, *OOH) valence states generate bonding/antibonding states, and the relative occupation of antibonding states governs the net bond strength. Within the Hammer-Nørskov framework [44], $\varepsilon_d$ is computed from the projected density of states (PDOS) as:

$$\epsilon_d = \frac{\int_{-\infty}^{+\infty} E.\rho(E)dE}{\int_{-\infty}^{+\infty} \rho(E)dE} \tag{4}$$

Here, $\rho(E)$ is the PDOS, and the integration is performed over −1.0 to 0 eV, considering only the occupied *d* orbitals near the Fermi level. As summarized in Table 3, the slight upward shift of $\varepsilon_d$ toward the Fermi level under +3% strain, indicating the enhanced coupling between Pd-*d* states and adsorbate frontier orbitals. This strain-induced upward shift redistributes the Pd states near the Fermi level, enhancing electronic coupling to the adsorbate frontier orbitals and leading to a corresponding modification of intermediate energy barriers. Notably, the adsorbate-dependent $\varepsilon_d$ values further reveal that strain affects different intermediates in a non-uniform manner. For HER, $\varepsilon_d$ shifts from −0.64 to −0.60 eV upon *H adsorption under +3% strain, suggesting enhanced Pd-H interaction, consistent with the reduced $\Delta G_{HER}$ barrier observed under tensile strain (Fig. 7). For OER intermediates, the response is intermediate-specific: $\varepsilon_d$ changes from −0.68 to −0.66 eV for *OH and from −0.56 to −0.59 eV for *OOH, while *O shifts from −0.74 to −0.78 eV, indicating preferential stabilization of the deprotonated oxygen intermediate under strain. This trend indicates that the strain modifies the Pd-Te and Pd-O hybridization, changing the energy alignment and occupation of Pd-adsorbate antibonding states. As a result, certain intermediates (notably *O) can be stabilized more strongly, while the energetic cost associated with O-O bond formation toward *OOH (*O to *OOH) increases, since this step is highly sensitive to the active site and occupation of adsorbate–metal antibonding states [45].

**Table 3**
Calculated d-band center for adsorbed intermediates on strained and unstrained penta-$PdTe_2$. Here * represents the penta-$PdTe_2$ monolayer.

| Surface absorbate | Unstrained monolayer (eV) | Monolayer with +3% strain (eV) |
|---|---|---|
| * | −0.57 | −0.55 |
| *H | −0.64 | −0.60 |
| *O | −0.74 | −0.78 |
| *OH | −0.68 | −0.66 |
| *OOH | −0.56 | −0.59 |

### 3.5. Solar to hydrogen efficiency

Solar-to-hydrogen (STH) efficiency is another key performance parameter in photocatalytic water splitting, quantifying a photocatalyst's ability to convert incident solar energy into chemical energy stored in hydrogen fuel. The parameter, $\eta_{abs}$, describes the fraction of incident solar radiation effectively absorbed by the photocatalyst, $\eta_{cu}$ denotes the efficiency with which photogenerated charge carriers are utilized in redox reactions, and $\eta_{STH}$ corresponds to the overall solar-to-hydrogen conversion efficiency. Under the idealized assumption of 100% catalytic efficiency, the theoretical upper limits of light absorption efficiency, carrier utilization efficiency, and the overall STH efficiency can be estimated. The solar light absorption efficiency is defined as:

$$\eta_{abs} = \frac{\int_{E_g}^{\infty} P(\hbar\omega)\, d(\hbar\omega)}{\int_{0}^{\infty} P(\hbar\omega)\, d(\hbar\omega)} \tag{5}$$

where $P(\hbar\omega)$ is the solar energy flux at a given photon energy, $\hbar\omega$, evaluated under AM 1.5G illumination, and $E_g$ is the bandgap energy of the photocatalyst. The carrier utilization efficiency is expressed as:

$$\eta_{cu} = \frac{\Delta G \int_{E}^{\infty} \frac{P(\hbar\omega)}{\hbar\omega} d(\hbar\omega)}{\int_{E_g}^{\infty} P(\hbar\omega) d(\hbar\omega)} \tag{6}$$

where $\Delta G = 1.23$ eV is the Gibbs free energy corresponding to the redox potential difference between the $H^{+}/H_2$ and $H_2\,O/O_2$ and $E$ is the energy of available photons. The integration from $E$ to $\infty$ in the numerator represents the effective photocurrent density generated by the absorbed solar spectrum.

$$E = \begin{cases} E_g, (U_e \geq 0.2, (U_h - 1.23) \geq 0.6) \\ E_g + 0.2 - U_e, (U_e < 0.2, (U_h - 1.23) \geq 0.6) \\ E_g + 0.6 - (U_h - 1.23), (U_e \geq 0.2, (U_h - 1.23) < 0.6) \\ E_g + 0.8 - U_e - (U_h - 1.23), (U_e < 0.2, (U_h - 1.23) < 0.6) \end{cases} \tag{7}$$

Finally, the total solar-to-hydrogen efficiency is then obtained as:

$$\eta_{STH} = \eta_{abs} \times \eta_{cu} \tag{8}$$

The effective energy term $E$, which accounts for kinetic losses due to overpotentials, can be evaluated explicitly in terms of the overpotentials associated with the HER and OER as follows:

$$E = \begin{cases} E_g, (U_e \geq 0.2, (U_h - 1.23) \geq 0.6) \\ E_g + 0.2 - U_e, (U_e < 0.2, (U_h - 1.23) \geq 0.6) \\ E_g + 0.6 - (U_h - 1.23), (U_e \geq 0.2, (U_h - 1.23) < 0.6) \\ E_g + 0.8 - U_e - (U_h - 1.23), (U_e < 0.2, (U_h - 1.23) < 0.6) \end{cases} \tag{9}$$

Here, $E_g$, $U_e$ and $U_h$ are the calculated band gap, photogenerated carriers (electrons and holes), respectively.

Table 4 summarizes the calculated available photon energy $E$, light absorption efficiency, carrier utilization efficiency, and the corresponding theoretical $\eta_{STH}$ for penta-$PdTe_2$ under unstrained and tensile-strain conditions, together with other pentagonal-based photocatalysts reported in the literature. For the unstrained monolayer ($\varepsilon_{xy} = 0\%$), $\eta_{STH}$ is predicted to be 8.46% at pH = 0 (with $E = 2.56$ eV, $\eta_{abs} = 42.67\%$, and $\eta_{cu} = 19.83\%$). At pH = 7, the required photon energy decreases to 2.15 eV, accompanied by increases in both light harvesting ($\eta_{abs} = 47.67\%$) and carrier utilization ($\eta_{cu} = 37.28\%$), leading to a markedly higher $\eta_{STH}$ of 15.91%. Upon applying +3% biaxial tensile strain, the available photon energy is further reduced to 2.32 eV (pH = 0) and 2.04 eV (pH = 7), while $\eta_{abs}$ remains high (∼47.60%) and $\eta_{cu}$ improves substantially (29.21% at pH = 0 and 42.86% at pH = 7). Consequently, the theoretical $\eta_{STH}$ increases to 13.91% at pH = 0 and reaches 20.40% at pH = 7, indicating that tensile strain significantly enhances overall energy conversion efficiency, primarily by improving carrier utilization. Notably, the strained penta-$PdTe_2$ at pH = 7 outperforms several reported pentagonal photocatalysts; such as penta-$PdSe_2$ (12.59%) [33], penta-$HgS_2$ (11.83%) [37], penta-NiSSe (11.25%) [46], penta-SiPAs (10.30%) [47], and penta-PdSeTe (12.08%) [48]; highlighting the strong potential of strain engineering in optimizing pentagonal 2D monolayers for solar-driven hydrogen production.

## 4. Conclusion

A comprehensive first-principles investigation of the photocatalytic water splitting properties of the novel pentagonal $PdTe_2$ monolayer is presented. Our aim was to investigate how catalytic efficiency can be improved by applying biaxial strain under bias in both neutral (pH = 7) and acidic (pH = 0) media. Special attention was focused on finding suitable conditions for the penta-$PdTe_2$ monolayer to accommodate both HER and OER processes. To assess the plausibility of their occurrences, we calculated the complete Gibbs free energy pathways (ΔG) at equilibrium potential (U = 0), applied bias (U = $U_e$), and beyond the rate-determining step (U > RDS) to estimate the overpotentials $\eta_{HER}$ and $\eta_{OER}$ corresponding to HER and OER, respectively, and leading to water splitting. As a requirement for spontaneous photocatalytic water splitting, the valence-band maximum (VBM) and the conduction-band maximum (CBM) were determined using the HSE06 functional and compared to the water redox potentials ($H^{+}/H_2$ at 0 V and $O_2/H_2O$ at 1.23 eV vs NHE). Finally, we evaluated the solar-to-hydrogen (STH) efficiency by integrating the optical absorption spectra with

**Table 4**
Calculated available photon energy (E), energy conversion efficiency of light absorption ($\eta_{abs}$), carrier utilization ($\eta_{cu}$), and theoretical STH ($\eta_{STH}$) of penta-$PdTe_2$ monolayer with and without biaxial strain ($\varepsilon_{xy}$).

| Material | pH | E (eV) | $\eta_{abs}$(%) | $\eta_{cu}$(%) | $\eta_{STH}$(%) | Reference |
|---|---|---|---|---|---|---|
| Penta-$PdTe_2$ ($\varepsilon_{xy} = 0\%$) | pH = 0 | 2.56 | 42.67 | 19.83 | 8.46 | This work |
| | pH = 7 | 2.15 | 47.67 | 37.28 | 15.91 | |
| Penta-$PdTe_2$ ($\varepsilon_{xy} = 3\%$) | pH = 0 | 2.32 | 47.60 | 29.21 | 13.91 | |
| | pH = 7 | 2.04 | 47.60 | 42.86 | 20.40 | |
| Penta-$PdSe_2$ | pH = 0 | – | 30.42 | 41.40 | 12.59 | [33] |
| Penta-$HgS_2$ | pH = 0 | – | 26.05 | 45.43 | 11.83 | [37] |
| Penta-NiSSe | pH = 0 | – | 24.97 | 45.05 | 11.25 | [46] |
| Penta-SiPAs | pH = 0 | – | 23.50 | 44.52 | 10.30 | [47] |
| Penta-PdSeTe | pH = 0 | – | 33.66 | 35.90 | 12.08 | [48] |

thermodynamic driving forces. The results can be summarized as follows:

1. Pristine pentagonal $PdTe_2$ monolayer exhibits paramagnetic semiconducting behavior with an energy bandgap of 1.87 eV. Under compressive strain, the band gap can be reduced.
2. The variation of VBM and CBM versus strain [−4%, +4%] with respect to vacuum level was compared to the redox level of water splitting in both neutral (pH = 7) and acidic (pH = 0) media. At tensile strains of +2% and +3%, the band edges encompass both the oxidation and reduction potentials at pH = 0 and 7.
3. Calculations of carrier mobility and effective mass have shown that the $PdTe_2$ monolayer is distinguished by high mobility and low effective mass of holes compared to many other pentagonal and hexagonal 2D materials. The high hole mobility facilitates rapid extraction of photogenerated holes to OER sites.
4. The Gibbs free energy results showed that the overpotentials were sensitive to tensile strain. At tensile strain of +3%, one can obtain the lowest overpotentials in both HER and OER processes with $\eta_{HER}$ = 0.70 V (at pH = 0), 1.60 V (at pH = 7) and $\eta_{OER}$ = 1.54 V (at pH = 0), 0.72 V (at pH = 7), respectively.
5. The solar to hydrogen efficiency (STH) increases to 13.91% at pH = 0 and reaches 20.40% at pH = 7, outperforming several pentagonal photocatalysts reported in literature. This reveals the strong potential of strain engineering in optimizing the penta-$PdTe_2$ monolayer for solar-driven hydrogen production.

The above findings propose pentagonal $PdTe_2$ as a promising candidate for efficient photocatalytic water splitting applications.

## CRediT authorship contribution statement

**Narender Kumar:** Writing – review & editing, Writing – original draft, Validation, Investigation, Formal analysis, Data curation. **Shambhu Bhandari:** Writing – review & editing, Investigation, Formal analysis, Data curation. **Dario Alfè:** Writing – review & editing, Investigation, Formal analysis, Data curation. **Ravindra Pandey:** Writing – review & editing, Supervision, Investigation, Conceptualization. **Nacir Tit:** Writing – original draft, Supervision, Investigation, Funding acquisition, Conceptualization.

## Declaration of competing interest

The authors declare that they have no known competing financial interests or personal relationships that could have appeared to influence the work reported in this paper.

## Acknowledgements

The authors are indebted to the support of the Research Office at the United Arab Emirates University under grant number 12R162. One of us (N.K.) would extend his thanks to the College of Graduate Studies (CGS) for granting him the PhD Fellowship.

## Appendix A. Supplementary data

Supplementary data to this article can be found online at https://doi.org/10.1016/j.apsusc.2026.167411.

## Data availability

Data will be made available on request.

**Narender Kumar** earned an MSc in Physics from Himachal Pradesh University, Summer Hill, Shimla, India. He is currently a PhD student in the Physics Department at the United Arab Emirates University, under the supervision of Prof. Nacir Tit and co-supervision of Prof. Ravindra Pandey (Michigan Technological University, USA). His research interests are in computational materials science, particularly DFT methods. As applications, he is interested in gas sensing, spintronics, energy harvesting and storage.

**Shambhu Bhandari Sharma** earned an MSc in Physics from Tribhuvan University, Kathmandu, Nepal. He is currently a PhD student in the Department of Earth Sciences at the University College London under the supervision of Prof. Dario Alfiè. His work extensively applies density functional theory, ab initio molecular dynamics and machine learning techniques to investigate material properties under high-pressure, high-temperature conditions comparable to those in planetary cores. He also has expertise in electronic structure calculations of low-dimensional materials and has authored several publications in his field.

**Dario Alfè** is a condensed-matter physicist who develops and applies first-principles computational tools to study the high-pressure and high-temperature properties of materials, primarily in the context of planetary interiors. Born and raised in Napoli, Italy, he received his Ph.D. from SISSA in Trieste, Italy. After post-doctoral stints at Keele University in England and University College London, in 2006, he joined the faculty of University College London. Since 2018, he has also been a faculty member of the Università di Napoli Federico II in Italy.

**Ravindra Pandey** did his PhD in Theoretical Solid-State Physics from the University of Manitoba, Canada. Currently, he is a professor of Physics at Michigan Tech University, USA. His research interests are in Computational Materials Physics. These include developing theoretical methods and computer programs, as well as analysing specific materials and defects. He is a Fellow of the American Physical Society.

**Nacir Tit** got both his MSc in Physics and PhD in Physics from the University of Minnesota at Minneapolis, USA, in 1988 and 1991, respectively. He worked for 2 years as a Postdoctoral Fellow at the Abdus Salam ICTP in Trieste, Italy. Thereafter, he joined the Physics Department at the UAE University, where he was promoted twice, becoming a Full Professor of Physics in 2003. Currently, he is a Professor of Physics at the UAE University. He is interested in research using DFT codes in the areas of photonics, gas sensing for the detection of toxic gases and VOCs related to cancer biomarkers, energy storage, metal-ion batteries, and spintronics.